\documentclass[10pt,conference]{IEEEtran}
\IEEEoverridecommandlockouts
\usepackage{cite}
\usepackage{amsmath,amssymb,amsfonts}
\usepackage{algorithmic}
\usepackage{hyperref}
\usepackage{graphicx}
\usepackage{textcomp}
\usepackage{xcolor}
\usepackage{xspace}
\usepackage{epsfig}
\usepackage{color,xcolor}
\usepackage{tcolorbox}
\usepackage{enumitem}
\tcbuselibrary{breakable}

\usepackage{booktabs}
\usepackage{tabularx}
\usepackage{colortbl}
\usepackage{makecell}
\usepackage{url}
\usepackage{tikz}
\usepackage{pgfplots}
\pgfplotsset{compat=1.18}
\usepackage{subfigure}
\usepackage{threeparttable}
\usepackage{booktabs}
\definecolor{lightgreen}{HTML}{E2F0D9}
\usepackage{makecell}
\usepackage{tabularx}
\usepackage{hhline} 
\usepackage{makecell} 

\def\BibTeX{{\rm B\kern-.05em{\sc i\kern-.025em b}\kern-.08em
    T\kern-.1667em\lower.7ex\hbox{E}\kern-.125emX}}
\begin{document}
\raggedbottom

\newcommand{\tool}{\textsc{AgentLens}\xspace}

\title{\tool: Graph-guided Security Auditing\\
for LLM-based Agents}

\author{%
    \IEEEauthorblockN{%
        Peizhuo~Lv\IEEEauthorrefmark{1}\IEEEauthorrefmark{4},
        Tao~Chen\IEEEauthorrefmark{2}\IEEEauthorrefmark{4},
        Mengya~Shi\IEEEauthorrefmark{2},
        Ziyu~Lin\IEEEauthorrefmark{1},
        Yuxuan~Chen\IEEEauthorrefmark{2}\IEEEauthorrefmark{5}, \\
        Ruigang~Liang\IEEEauthorrefmark{3}\IEEEauthorrefmark{5},
        Tianwei~Zhang\IEEEauthorrefmark{1},
        Wei~Dong\IEEEauthorrefmark{1}\IEEEauthorrefmark{5},
        XiaoFeng~Wang\IEEEauthorrefmark{1}%
    }
    \IEEEauthorblockA{\IEEEauthorrefmark{1}Nanyang Technological University, Singapore}
    \IEEEauthorblockA{\IEEEauthorrefmark{2}School of Cyber Science and Technology, Shandong University, China}
    \IEEEauthorblockA{\IEEEauthorrefmark{3}Institute of Information Engineering, Chinese Academy of Sciences, China}
    \IEEEauthorblockA{%
        Email: peizhuo.lyu@ntu.edu.sg, tchen200311@gmail.com, mengyashi752@gmail.com, \\
        linziyu0205@gmail.com, chenyuxuan@sdu.edu.cn, liangruigang@iie.ac.cn, \\
        tianwei.zhang@ntu.edu.sg, wei\_dong@ntu.edu.sg, xiaofeng.wang@ntu.edu.sg%
    }
    \thanks{
        \IEEEauthorrefmark{4}Equal contribution.
        \quad\IEEEauthorrefmark{5}Corresponding authors.%
    }
}

\maketitle

\begin{abstract}

LLM-based agents are emerging as a key paradigm for automating software engineering by coupling LLM-driven decision-making with external tools, files, APIs, and execution environments. However, weaknesses in agent implementations and insecure interactions between the agent and external resources can lead to severe consequences, including remote code execution, arbitrary file access, SQL injection, and data leakage. This makes pre-deployment auditing essential for systematically identifying implementation weaknesses and validating whether they can manifest as concrete security risks in realistic agent workflows. Existing approaches either focus on specific vulnerability classes or lack the capability to validate whether identified static weaknesses correspond to executable exploits.

In this paper, we propose \tool, a systematic auditing framework for LLM-based agents. \tool analyzes the target agent’s implementation, component interactions, and built-in defenses, and organizes them into an \textit{Audit Knowledge Graph} that captures how these factors interact to expose potential security risks. Based on this graph, \tool derives candidate  attack paths, generates path-specific validation payloads, and constructs realistic multi-turn audit scenarios that place these payloads in task-relevant contexts to validate whether suspected risks manifest as security-violating behaviors during execution.
Across five widely-used agent applications, each with over 24,000 GitHub stars, \tool identifies 44 vulnerabilities, and every vulnerability is validated by being triggered during execution.

Compared with the state-of-the-art approach (AgentFuzz), \tool detects 3.67$\times$ more vulnerabilities. To date, 12 reported issues have been assigned CVE IDs.

\end{abstract}

\section{Introduction}

LLM-based agents have become prevalent in software engineering workflows, including code generation~\cite{Wang2024TowardsUT}, program repair~\cite{Bouzenia2024RepairAgentAA}, repository understanding~\cite{Ahmad2026OpenSWETracesAD}, test generation~\cite{Chen2023ChatUniTestAF}, and security testing~\cite{Ding2024VulnerabilityDW}. They have been integrated into IDE extensions~\cite{Bajpai2024LetsFT}, command-line development assistants~\cite{Yang2024SWEagentAI}, cloud coding platforms~\cite{Bhatnagar2025ACF}, automated maintenance workflows~\cite{Bruches2026TAMEvalEL}, and web-facing software services~\cite{Pedro2023FromPI}.
In such settings, agent decisions are no longer confined to textual responses; they can directly drive tool invocations, file operations, API calls, and other state-changing actions. 

The growing deployment of LLM-based agents in diverse real-world scenarios, together with their expanding access to tools, memory, data, and external services, makes pre-deployment security auditing essential. Particularly, prior studies~\cite{Liu2023DemystifyingRV, Pedro2023FromPI, Yang2026UnderstandingAM} on LLM-integrated applications have shown that insecure coupling between LLM-generated content and executable operations can lead to severe consequences, including remote code execution, arbitrary file access, SQL injection, and prompt leakage. These risks become more pronounced when agents iteratively observe context, plan actions, invoke tools, and update external states during task execution. Therefore, auditing is needed to identify implementation weaknesses and validate whether they can lead to security-violating behaviors in realistic workflows, thereby helping developers prioritize subsequent security hardening.

Prior security testing approaches for foundation LLMs~\cite{Zou2023UniversalAT} are insufficient for auditing agents. LLM-centric red-teaming approaches (HarmBench~\cite{Mazeika2024HarmBenchAS}, garak~\cite{Derczynski2024garakAF}, and PyRIT~\cite{csa2024pyrit}) primarily evaluate whether adversarial prompts can elicit unsafe or policy-violating responses from a target LLM. However, agent security depends not only on the LLM response itself, but also on how the surrounding implementation interprets that response and translates it into tool calls, external requests, or other environment-changing actions. As a result, a semantically benign LLM output may still lead to an insecure operation, whereas a harmful LLM output may be blocked by guardrails before affecting the environment.
Recent studies have examined specific vulnerability classes in LLM-integrated applications, which are closer to agent auditing. For example, AgentFuzz~\cite{Liu2025MakeAD} detects taint-style vulnerabilities in agent systems, while Prompt-to-SQL injection~\cite{Pedro2023FromPI} studies how prompts can induce insecure database queries.
Different from them, agent auditing should focus on how to comprehensively assess the security of an LLM agent, rather than only detecting a predefined class of vulnerabilities. 
Particularly, for a complex agent, the security risks are often unknown beforehand: where they arise, what types they belong to, which component interactions cause them, whether they are affected by existing defenses, and how they should be tested. 

Recently, Agent Audit~\cite{Zhang2026AgentAA} was proposed to audit agents by statically analyzing their code and deployment artifacts to detect risky tool implementations, insecure data flows, etc. However, vulnerability candidates found by static analysis do not guarantee that the reported issues can be triggered in an actual agent execution. Specifically, Agent Audit identifies no validated vulnerabilities in Langflow~\cite{langflow2026} (as shown in Sec.~\ref{sec:comparison}). This is because static approaches lack dynamic context-awareness: they cannot determine whether a vulnerable tool is reachable under the LLM's non-deterministic task planning, nor can they assess whether built-in dynamic guardrails (e.g., intent filters, permission
checks) would block the execution. This shows that effective agent auditing must account for not only the suspicious implementation weaknesses, but also whether the reported issues are feasible and can lead to security-violating outcomes within the agent's executable workflow. 

We highlight two technical fundamental challenges that need to be addressed for agent auditing: (C1) \textit{How to define the audit scope of a target agent?}
Agent risks are often implicit and relational: they arise from interactions among prompts, tools, files, APIs, configurations, execution environments, and built-in defenses, rather than isolated implementation flaws. Thus, the auditor must infer the target agent’s security-sensitive behaviors and formulate concrete audit objectives. Currently, the security boundary of an agent is not well defined. 
(C2) \textit{How to validate whether a suspected risk manifests in realistic agent workflows?}
A suspicious weakness is not necessarily exploitable: it may be unreachable from attacker-controlled inputs, depend on specific agent states, or be blocked by intermediate defenses. The auditor must therefore turn each suspected risk into an executable scenario and collect reliable evidence that it leads to a security violation.

To address these challenges, we propose \tool, a systematic auditing framework for LLM-based agents. The core of \tool is a new data structure, \textit{Audit Knowledge Graph}, to model the target agent's key components, dependencies, execution relations, potential weaknesses, and built-in guardrails. This graph allows \tool to derive concrete audit objectives from the target agent itself, rather than assuming a predefined vulnerability class (addressing C1). Besides, it also guides the generation of candidate audit paths and the corresponding payloads, which will be validated for the potential of security violation (addressing C2). 

Specifically, \tool first analyzes the target agent's code and configuration files to construct an Agent Profile Graph, whose nodes represent agent components and edges represent invocation, data-flow, and control-flow dependencies. It then enriches this graph with potential weaknesses associated with component nodes and deployed guardrails represented as constraints, forming an \textit{Audit Knowledge Graph}. Based on this graph, \tool derives candidate audit paths that connect external agent entry points with security-critical operations while preserving the associated weakness evidence and defense constraints. For each path, \tool uses the retained path context to generate a validation payload to drive the agent toward the corresponding security-critical behavior under its original defenses. Because many security-critical operations are invoked only within specific task contexts rather than by standalone interactions, \tool further constructs multi-turn audit scenarios that guide the agent through the relevant workflow and deliver the payload at the appropriate execution point. 
Finally, \tool checks whether the actual execution results show a concrete security violation.

We evaluate \tool on five widely used open-source software engineering agents: Aider, MetaGPT, Goose, Cline, and Roo-Code. 
\tool identifies 44 validated vulnerabilities across multiple categories, including Code Injection, Path Traversal, and Server-Side Request Forgery.
These vulnerabilities lead to diverse security impacts, including credential access, system compromise, code tampering, and data exfiltration, etc.
To date, 12 reported issues have been assigned CVE IDs: CVE-2026-10174 to CVE-2026-10177, CVE-2026-19058 to CVE-2026-19060, and CVE-2026-81833 to CVE-2026-81837.
\tool takes 2,396.44 seconds per agent on average for the auditing workflow.
It generates an average of 14 audit paths and identifies an average of 8.8 validated vulnerabilities per agent.
The ratio of validated vulnerabilities to audit paths is 62.86\%.
Compared with existing agent security testing techniques on LangFlow, \tool detects 11 vulnerabilities, whereas AgentFuzz and Agent Audit detect 3 and 0, respectively.
Our ablation study further shows that the core components contribute to vulnerability discovery.
Finally, we present a case study of server-side request forgery in Aider.
\tool connects isolated weaknesses into a concrete audit path.
It then confirms the vulnerability under Aider's original workflow and guardrails.

\vspace{2pt}\noindent\textbf{Contributions.} We summarize our contributions as follows:

\begin{itemize}[left=0em]
\item We propose \tool, a systematic auditing framework for LLM-based agents. \tool builds an Audit Knowledge Graph that captures agent workflow components and dependencies, weakness evidence, and guardrail constraints, allowing agent auditing to be performed through a path-oriented analysis problem.

\item We design an automated validation approach that generates audit payloads and constructs multi-turn audit scenarios from the derived audit paths. This approach enables \tool to validate whether isolated weaknesses can propagate through the agent's original workflow and manifest as security-violating behaviors under existing guardrails.

\item We comprehensively evaluate \tool on five widely used open-source agents.
\tool identifies 44 validated vulnerabilities, 12 of which have been assigned CVE IDs.
We provide our implementation in an anonymized repository to facilitate LLM-based agent security auditing.
\end{itemize}

\section{Overview}

\subsection{Threat Model}
\label{sec:threat-model}

We consider a pre-deployment auditing scenario for LLM-based agents. The auditor is a trusted party, typically a maintainer or in-house security engineer, who evaluates the target agent prior to production deployment, and the auditing environment is assumed to be uncompromised. The auditor aims to discover implementation-level weaknesses and validate whether they can manifest through the agent's normal workflows as security-violating behaviors.
Following common risk categories discussed in OWASP Top 10 for Agentic Application~\cite{owasp2026agentic}, we consider common security violations such as tool misuse, unexpected code execution, and privilege abuse.
We consider sensitive agent actions arising from both direct interactions and untrusted external inputs.
In multi-user deployments, malicious users may affect other benign users through shared agent resources or workflows.
We assume the auditor has access to the target agent's implementation artifacts, including source code, configuration files, prompts, tool definitions, API bindings, permission settings, and guardrail mechanisms. The auditor can use external threat intelligence~\cite{nistnvd2026, githubadvisories2026, snykvulndb2026, hackeronehacktivity2026, zdiadvisories2026} and automated weakness discovery tools~\cite{semgrep2026, codeql2026, praetorianFreeBSDKernel2026} to identify potential weaknesses in agent components and their dependencies. The auditor can also execute the agent in a controlled testing environment that preserves its original workflows and defenses, interact with the agent through its legitimate external entry points, and observe execution outputs, tool invocations, logs, and state changes to validate suspected risks under the agent’s original execution.

\subsection{Design Insight }
\label{subsec:philosophy}

Existing systems such as Agent Audit~\cite{Zhang2026AgentAA} can identify isolated security-relevant weaknesses in the implementation. However, this is insufficient for agent auditing. An LLM-based agent converts user instructions into executable actions through multi-step workflows involving reasoning, tool invocation, resource access, and code or API execution. High-impact security risks, such as remote code execution or SQL injection, often become reachable after the agent follows a task-specific workflow, rather than through a single execution step. A reported weakness can be validated as a real risk only when it is reachable from an external entry point and can propagate through the agent workflow to a security-critical operation under the agent's built-in guardrail mechanisms. Thus, agent auditing needs to \textbf{analyze the relations within an agent workflow}, such as how components interact, whether a weakness can be reached and triggered, how multiple invocation steps lead to a security-critical operation, and whether the execution can proceed under the agent's built-in guardrails. 

Driven by the above auditing demand, our key idea is to model the multi-step agent workflow as a graph. The graph provides a natural abstraction for representing entities, their dependencies, and reachability, thereby making the execution relations explicit and analyzable. We introduce a novel graph structure, \textit{Audit Knowledge Graph}, to represent key agent components as nodes, while invocation, data-flow, and control-flow dependencies as edges. Beyond agent workflow connectivity, Audit Knowledge Graph annotates component nodes with potential weaknesses identified through threat intelligence collection or weakness discovery tools, while encoding built-in guardrails as constraints that affect the path feasibility.

With this relational structure, \textbf{agent auditing can be formulated as a path-oriented analysis problem over the Audit Knowledge Graph}: candidate audit paths can be derived from external entry nodes to security-critical operation nodes while accounting for how relevant weaknesses may be triggered and how guardrails along the path may be satisfied or bypassed.
These paths provide the workflow context needed to generate validation payloads that target the security-critical operations. When the security-critical operation is reachable only through a multi-step workflow, such paths also guide the construction of multi-turn audit scenarios that deliver the payload to the intended operation.

\begin{figure}
\centering
\begingroup
\definecolor{akgWhite}{HTML}{FFFFFF}
\definecolor{akgBlue}{HTML}{2B5E86}
\definecolor{akgInk}{HTML}{20313E}
\definecolor{akgBorder}{HTML}{BAC6CF}
\definecolor{akgMuted}{HTML}{6D7B87}
\definecolor{akgPale}{HTML}{F5F8FA}
\definecolor{akgRisk}{HTML}{9A5B23}
\definecolor{akgGuard}{HTML}{2E6B4A}
\definecolor{akgGuardFill}{HTML}{EFF6F0}
\definecolor{akgRiskFill}{HTML}{FFF4E8}
\resizebox{\columnwidth}{!}{%
\begin{tikzpicture}[x=1pt,y=-1pt,line width=0.55pt,
 every node/.style={inner sep=0pt,outer sep=0pt,anchor=base west,text=akgInk}]
\path[use as bounding box] (0,0) rectangle (252,216);
\draw[draw=akgBorder,fill=akgPale,rounded corners=3pt,line width=0.55pt] (0.5,0.5) rectangle (251.5,97.0);
\node[anchor=base west,text=akgInk,font=\bfseries\fontsize{7.6pt}{8.9pt}\selectfont] at (8,11) {Original code};
\draw[draw=akgBorder,line width=0.55pt] (8,15) -- (244,15);
\node[anchor=base west,text=akgMuted,font=\fontsize{6.5pt}{7.8pt}\selectfont] at (7,25) {1};
\node[anchor=base west,text=akgBlue,font=\ttfamily\fontsize{7.2pt}{8.5pt}\selectfont] at (21.0,25) {\detokenize{def reply_completed(self):}};
\node[anchor=base west,text=akgMuted,font=\fontsize{6.5pt}{7.8pt}\selectfont] at (7,33) {2};
\node[anchor=base west,text=akgInk,font=\ttfamily\fontsize{7.2pt}{8.5pt}\selectfont] at (36.12,33) {\detokenize{content = self.partial_response_content}};
\fill[akgGuardFill] (18,34.2) rectangle (247,42.8);
\node[anchor=base west,text=akgMuted,font=\fontsize{6.5pt}{7.8pt}\selectfont] at (7,41) {3};
\node[anchor=base west,text=akgInk,font=\ttfamily\fontsize{7.2pt}{8.5pt}\selectfont] at (36.12,41) {\detokenize{if not content or not content.strip():}};
\node[anchor=base west,text=akgMuted,font=\fontsize{6.5pt}{7.8pt}\selectfont] at (7,49) {4};
\node[anchor=base west,text=akgInk,font=\ttfamily\fontsize{7.2pt}{8.5pt}\selectfont] at (51.239999999999995,49) {\detokenize{return}};
\fill[akgGuardFill] (18,50.2) rectangle (247,58.8);
\node[anchor=base west,text=akgMuted,font=\fontsize{6.5pt}{7.8pt}\selectfont] at (7,57) {5};
\node[anchor=base west,text=akgInk,font=\ttfamily\fontsize{7.2pt}{8.5pt}\selectfont] at (36.12,57) {\detokenize{if not self.io.confirm_ask("Edit the files?"):}};
\node[anchor=base west,text=akgMuted,font=\fontsize{6.5pt}{7.8pt}\selectfont] at (7,65) {6};
\node[anchor=base west,text=akgInk,font=\ttfamily\fontsize{7.2pt}{8.5pt}\selectfont] at (51.239999999999995,65) {\detokenize{return}};
\node[anchor=base west,text=akgMuted,font=\fontsize{6.5pt}{7.8pt}\selectfont] at (7,73) {7};
\node[anchor=base west,text=akgInk,font=\ttfamily\fontsize{7.2pt}{8.5pt}\selectfont] at (36.12,73) {\detokenize{editor_coder = Coder.create(**kwargs)}};
\fill[akgRiskFill] (18,74.2) rectangle (247,82.8);
\node[anchor=base west,text=akgMuted,font=\fontsize{6.5pt}{7.8pt}\selectfont] at (7,81) {8};
\node[anchor=base west,text=akgInk,font=\ttfamily\fontsize{7.2pt}{8.5pt}\selectfont] at (36.12,81) {\detokenize{editor_coder.run(with_message=content,}};
\fill[akgRiskFill] (18,82.2) rectangle (247,90.8);
\node[anchor=base west,text=akgRisk,font=\ttfamily\fontsize{7.2pt}{8.5pt}\selectfont] at (51.239999999999995,89) {\textcolor{akgRisk}{\detokenize{preproc=False}}\detokenize{)}};
\node[anchor=base west,text=akgInk,font=\bfseries\fontsize{7.5pt}{8.8pt}\selectfont] at (1,116) {Audit Knowledge Graph};
\draw[draw=akgBlue,fill=akgPale,rounded corners=3pt,line width=0.8pt] (0.5,128) rectangle (100.0,148);
\node[anchor=base,text=akgBlue,font=\ttfamily\fontsize{7.2pt}{8.5pt}\selectfont] at (50.25,140.5) {\detokenize{reply_completed()}};
\draw[draw=akgBorder,fill=akgWhite,rounded corners=3pt,line width=0.55pt] (0.5,182) rectangle (100.0,202);
\node[anchor=base,text=akgInk,font=\ttfamily\fontsize{7.2pt}{8.5pt}\selectfont] at (50.25,194.5) {\detokenize{editor_coder.run()}};
\draw[->,draw=akgBlue,line width=0.75pt] (27,148) -- (27,181);
\node[anchor=base west,text=akgInk,font=\fontsize{6.8pt}{8.1pt}\selectfont] at (33,162) {Invocation};
\node[anchor=base west,text=akgInk,font=\fontsize{6.8pt}{8.1pt}\selectfont] at (33,171) {dependencies};
\draw[dashed,draw=akgBlue,line width=0.6pt] (100,138) -- (109.5,138);
\draw[draw=akgBorder,fill=akgWhite,rounded corners=3pt,line width=0.55pt] (109.5,108) rectangle (251.5,215.5);
\node[anchor=base west,text=akgBlue,font=\fontsize{7.2pt}{8.5pt}\selectfont] at (115,119) {\textbf{Node\_name:} \texttt{reply\_completed()}};
\draw[draw=akgBorder,line width=0.55pt] (115,124) -- (246,124);
\node[anchor=base west,text=akgRisk,font=\bfseries\fontsize{7.3pt}{8.6pt}\selectfont] at (115,135) {Weakness:};
\node[anchor=base west,text=akgInk,font=\fontsize{7.2pt}{8.5pt}\selectfont] at (115,144) {$c$: CWE-693};
\node[anchor=base west,text=akgInk,font=\fontsize{7.2pt}{8.5pt}\selectfont] at (115,153) {$d$: \texttt{preproc=False} bypasses validation};
\node[anchor=base west,text=akgGuard,font=\bfseries\fontsize{7.3pt}{8.6pt}\selectfont] at (115,167) {Guardrails:};
\node[anchor=base west,text=akgInk,font=\fontsize{7.2pt}{8.5pt}\selectfont] at (115,176) {$m_1$: Empty check};
\node[anchor=base west,text=akgInk,font=\fontsize{7.2pt}{8.5pt}\selectfont] at (115,185) {$r_1$: Must be non-empty and non-whitespace};
\node[anchor=base west,text=akgInk,font=\fontsize{7.2pt}{8.5pt}\selectfont] at (115,197) {$m_2$: Human approval};
\node[anchor=base west,text=akgInk,font=\fontsize{7.2pt}{8.5pt}\selectfont] at (115,206) {$r_2$: User must confirm ``Edit the files?''};
\end{tikzpicture}%
}
\endgroup
\caption{An Example of Audit Knowledge Graph.}
\vspace{-8pt}
\label{fig:graph-case}
\vspace{-8pt}
\end{figure}
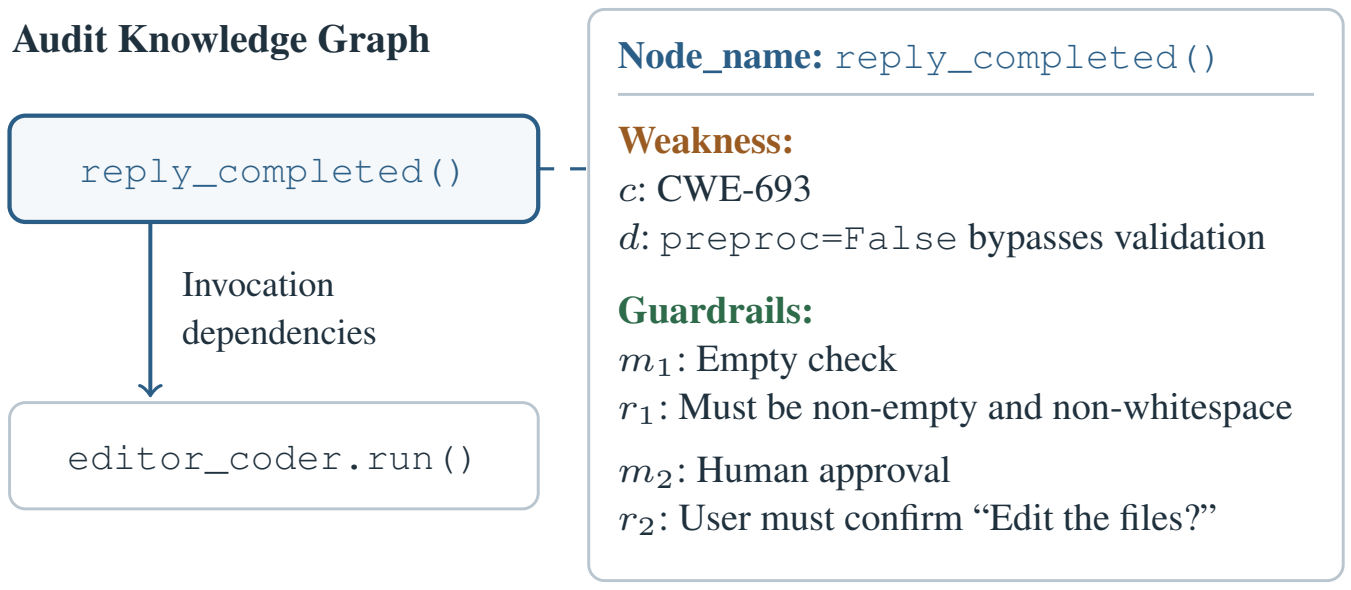

Figure~\ref{fig:graph-case} shows one example of constructing Audit Knowledge Graph from an agent implementation and identifying the exploitable risks. We consider a task in Aider's Architect mode~\cite{aider2026}, where Architect LLM passes code modification plans to the Editor LLM through function calls. We model these functions as nodes (i.e., \texttt{reply\_completed} and \texttt{editor\_coder.run()}) and their relation as ``Invocation dependencies'' edges. For \texttt{reply\_completed} node, we identify a CWE-693~\cite{cwe693} weakness by external threat intelligence because  \texttt{preproc=False} argument will disable the protection mechanism that validates Architect LLM's output before Editor Coder execution. Furthermore, we extract built-in guardrails (Lines 3--6). These guardrails are formatted into defense mechanisms (`m') and constraint rules (`r').

\section{Detailed Methodology}

\label{sec:approach}

\subsection{Overview}

\begin{figure*} 
\centering
\epsfig{figure=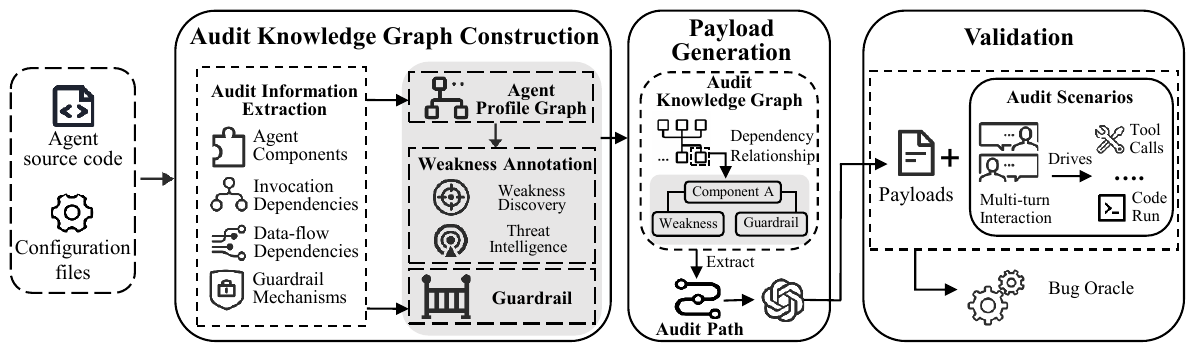, width=0.95\textwidth} 
\vspace{-5pt}
\caption{Workflow of Our \tool.}
\label{fig:workflow}
\vspace{-15pt}
\end{figure*}

We design \tool, a new framework for security auditing of LLM-based agents. Figure \ref{fig:workflow} shows the overview of its workflow, which consists of three concrete stages. (1) \textbf{Audit Knowledge Graph Construction} (Sec. \ref{sec:audit-graph}). \tool analyzes the source code and configuration files of the target agent to extract its key components, invocation dependencies, and data-flow dependencies. The extracted components and dependencies form an Agent Profile Graph. \tool then identifies weaknesses associated with these components and dependencies, and maps them back to the corresponding graph elements. In parallel, \tool extracts guardrail mechanisms from the agent implementation and configuration, and represents them as guardrail constraints to complete the Audit Knowledge Graph. (2) \textbf{Payload Generation} (Sec. \ref{sec:payload-generation}). Next, \tool generates audit payloads from the Audit Knowledge Graph. Starting from agent entry points, \tool derives candidate audit paths to security-critical operations that may be reached through the identified weaknesses. For each retained path, \tool synthesizes a payload by jointly considering the involved components, their associated weaknesses, and the guardrail constraints that must be satisfied or bypassed. (3) \textbf{Validation} (Sec. \ref{sec:validation}). 
Finally, \tool validates the generated payloads through scenario-based testing.
To match the multi-turn workflows of LLM-based agents, \tool embeds raw payloads into contextualized multi-turn audit scenarios.
\tool then checks whether the execution results show a concrete security violation and reports validated vulnerabilities.

\subsection{Audit Knowledge Graph Construction}
\label{sec:audit-graph}
In this stage, \tool first derives an Agent Profile Graph from the target agent's codebase via static analysis. It then constructs an Audit Knowledge Graph by augmenting the profile graph with weakness annotations obtained from threat intelligence collection and weakness discovery, together with guardrail constraints extracted from the target agent.

\subsubsection{Agent Profile Extraction} 
The Agent Profile Graph captures the execution structure of the target agent, including its components and dependencies among them. It is defined as:
\begin{equation}
\label{eq:agent-profile}
G_{p} = (N_{p}, E_{p}), 
\end{equation}
where $N_p$ denotes agent component nodes and $E_p$ denotes dependency edges among them. Each node $n_i\in N_p$ is represented as $n_i=\langle e_i,\tau_i,\phi_i\rangle$, where $e_i$ is the component entity, such as a specific class or function name in the Agent. $\tau_i$ is its component type that defines the functional category within the agent architecture (e.g., LLM interface, tool wrapper, or memory operation), and $\phi_i$ stores its associated attributes, including code locations and parameter lists.

To construct $G_p$, \tool parses the source code and configuration files of the target agent. It identifies class and function definitions using abstract syntax tree analysis and extracts LLM interfaces, external tools, and memory modules as component nodes. Furthermore, \tool analyzes the structural and behavioral features such as inheritance relationships, tool registration bindings, and external API calls. It then captures the invocation, data-flow, and control-flow relationships among these components as edges, and finally generates the Agent Profile Graph.

\subsubsection{Audit Knowledge Graph} 
After extracting Agent Profile Graph, we enrich it with weakness annotations and guardrail information to construct Audit Knowledge Graph. The graph captures how weaknesses in components and inter-component dependencies can be chained together to lead to security-critical behaviors. We define Audit Knowledge Graph as:
\begin{equation}
\label{eq:audit-graph}
G_{a} = (N_{p}, E_{p}) \cup (\mathcal{W}, \mathcal{G}).
\end{equation}
Here, $N_{p}$ and $E_{p}$ inherit the definitions from the Agent Profile Graph. The set $\mathcal{W}$ captures weaknesses identified from two complementary sources: (i) external threat intelligence collected from public agent issues~\cite{aiderissues2026, goosesecurity2026}, public vulnerability databases~\cite{nistnvd2026, githubadvisories2026, snykvulndb2026}, and security reports~\cite{hackeronehacktivity2026, zdiadvisories2026}; and (ii) implementation-specific weakness discovery over the target agent, which can be performed using automated weakness discovery tools such as Semgrep~\cite{semgrep2026} and CodeQL~\cite{codeql2026}. The set $\mathcal{G}$ records guardrails extracted from the agent's configurations, prompts, and defensive logic,  which are later used as defense constraints during payload generation.

\noindent\textbf{Weakness Annotation.}
We define the weakness set as $\mathcal{W}=\{w_{i}\}_{i=1}^{m}$. Each annotation is defined as $w_{i} = \langle e_{i}, c_{i}, d_{i} \rangle$, consisting of three factors: 
(i) \textit{target object} $e_{i}$, which identifies the component entity to which the weakness is attached and maps the annotation to the corresponding node in the Agent Profile Graph;
(ii) \textit{weakness category} $c_{i}$, which specifies the agent-related security risk associated with $e_{i}$. These categories are informed by CWE~\cite{mitrecwe2026};
(iii) \textit{weakness details} $d_{i}$, which record the risk descriptions.

\noindent$\bullet$ \textit{Threat Intelligence Collection}.
To collect external threat intelligence, \tool automatically queries and retrieves security reports from multiple public sources, including vulnerability databases (e.g., NVD~\cite{nistnvd2026}, GitHub Advisory Database~\cite{githubadvisories2026}, the Snyk Security Database~\cite{snykvulndb2026}), vulnerability disclosure platforms (e.g., HackerOne~\cite{hackeronehacktivity2026}, Zero Day Initiative~\cite{zdiadvisories2026}), as well as public issue reports and official security advisories~\cite{aiderissues2026, goosesecurity2026}. Specifically, we use the component entity $e_{i}$ and its associated type $\tau_{i}$, extracted from the Agent Profile Graph, as search keywords to query for relevant security reports. We extract the reported security risk as the weakness category $c_{i}$, and parse report details to record attack descriptions as the weakness details $d_{i}$. Finally, this threat intelligence is annotated onto the corresponding node $e_{i}$.

\noindent$\bullet$ \textit{Weakness Discovery}. 
In addition to identifying known vulnerabilities reported in other agents through external threat intelligence, \tool also discovers implementation-specific weaknesses within the target agent. To achieve this, we instantiate 10 automated discovery skills based on the critical security risks outlined in the OWASP Top 10 for Agentic Applications~\cite{owasp2026agentic}, and utilize Claude Code to apply these skills to scan the source code of the target agent. Specifically, each skill consists of the following factors: 
(i) \textit{target scope}, which specifies the classes, functions, or core agent modules to be analyzed. 
(ii) \textit{detection rules}, which provide the static patterns and regular expressions to identify suspicious code segments. 
(iii) \textit{logic constraints}, which define the necessary contextual conditions to confirm the weakness. 
(iv) \textit{output formats}, which standardize how to record the identified findings. 
For each finding, we extract the detected security risk as the weakness category $c_{i}$, record the risk descriptions as the weakness details $d_{i}$, and annotate them onto the corresponding node $e_{i}$. 

\noindent\textbf{Security Guardrails.} 
\tool extracts security guardrails from the target agent to characterize the defense constraints that may affect the attack feasibility. We denote the guardrail set as $G = \{g_i\}_{i=1}^{r}$, where each guardrail is represented as $g_i=\langle e_i,m_i,r_i\rangle$. Here, $e_{i}$ identifies the protected component entity and maps the guardrail to the corresponding node; $m_i$ denotes the defense mechanism, such as input validation, commit verification, or error handling; and $r_i$ denotes the concrete constraint rule, such as repository count limits or explicit pre-commit options. \tool extracts these guardrails from prompts, configuration files, tool wrappers, permission checks, and human-approval logic, and attaches them to the Audit Knowledge Graph.

\subsection{Payload Generation}
\label{sec:payload-generation}

Given the Audit Knowledge Graph, \tool derives candidate audit paths from entry nodes to security-critical operation nodes.
It then employs an LLM to generate test payloads guided by these paths.

\subsubsection{Audit Path Generation}
\tool traverses the Audit Knowledge Graph to derive candidate audit paths. Starting from attacker-controllable entry nodes, \tool performs depth-first traversal toward security-critical operation nodes and obtains an initial candidate path set $\mathcal{P}_{cand}$. Specifically, each path $p \in \mathcal{P}_{cand}$ is defined as a connected sequence of nodes $p = (N_{entry},  \dots, N_{crit})$. Here, $N_{entry}$ denotes an external interaction interface of the agent, such as a user prompt, uploaded file, retrieval query, or API request, through which untrusted content may enter the agent workflow. $N_{crit}$ denotes a security-critical component or operation, such as a code interpreter, system-command tool, file-system operation, database query, external API call, or privileged tool invocation. 

The initial candidate set $\mathcal{P}_{cand}$ often contains redundant paths.
This redundancy occurs because the agent's iterative reasoning and repeated tool calls generate many overlapping sub-paths and paths that share the same exploitation logic.
To filter out such redundancy, we apply a two-step reduction mechanism.
First, we remove redundant sub-paths that are completely contained within other paths.
Second, we group paths that share the same entry node, security-critical node, and weaknesses, keeping only one path from each group.
After filtering, we calculate a priority score for each remaining candidate path to determine the execution order for payload generation.
The priority score of a path $p$ is defined as
\begin{equation}
Priority(p) = \alpha \cdot \widehat{\mathcal{V}}(p)
- \beta \cdot \widehat{\text{Len}}(p).
\end{equation}
Here, $\mathcal{V}(p)$ represents the aggregated weakness severity score, and $\text{Len}(p)$ represents the path length.
Before weighting, we normalize $\mathcal{V}(p)$ and $\text{Len}(p)$ over the retained candidate path set $\mathcal{P}$:

\begin{equation}
\begin{aligned}
\widehat{\mathcal{V}}(p)
&= \frac{\mathcal{V}(p)}
{\max_{q \in \mathcal{P}} \mathcal{V}(q)}, \\
\widehat{\text{Len}}(p)
&= \frac{\text{Len}(p)}
{\max_{q \in \mathcal{P}} \text{Len}(q)}.
\end{aligned}
\end{equation}
If a denominator is zero, the corresponding normalized value is set to zero.
$\alpha$ and $\beta$ are positive hyper-parameters that adjust their weights\footnote{
In our implementation, we empirically set $\alpha=0.7$ and $\beta=0.3$.
}.
We measure the priority of a candidate path based on these two factors.
For the weakness severity, $\mathcal{V}(p)$ is calculated by summing the standardized scores for all weaknesses on the path.
A larger $\mathcal{V}(p)$ means the path contains more critical weaknesses, which increases the priority.
For the path length, a larger $\text{Len}(p)$ means the path involves more intermediate components.
While a longer path may expose more weaknesses, it introduces stricter execution constraints and makes the target harder to reach.
Thus, we treat $\text{Len}(p)$ as a term to reflect the execution difficulty.
A higher priority score places a candidate path earlier in the evaluation order.
When the audit budget is sufficient, the auditor can ignore the priority score and evaluate all retained candidate paths.\footnote{To identify as many potential security risks as possible, we tested all retained paths in our experiments.}
With a limited budget, the auditor evaluates higher-priority paths first.

\subsubsection{Test Payload Construction}
Given the selected audit paths, \tool employs an LLM to generate test payloads. Specifically, we provide the LLM with the following information: (1) risk category $\mathcal{R}$, which specifies the agent-related security risk, such as prompt injection or tool abuse. We use the OWASP Top 10 for Agentic Applications~\cite{owasp2026agentic} and CWE~\cite{mitrecwe2026} as a reference when defining these categories; (2) security impact $\mathcal{I}$, which specifies the expected consequence of a security violation, such as arbitrary code execution or data leakage, based on the MITRE ATLAS framework~\cite{mitreatlas2026}; (3) path information $p$, which describes the selected audit path, including the invoked intermediate components, their associated weaknesses, and the corresponding security guardrails. The above information guides the LLM to generate payloads grounded in the target agent's workflow and guardrail context. Specifically, we formalize this generation process as follows:
\begin{equation}
\label{equ:payload_gen}
Payload = \text{LLM}_{gen}(\mathcal{R}, \mathcal{I}, p)
\end{equation}
where $\text{LLM}_{gen}$ is configured with the following prompt. 
The design of this prompt considers that direct malicious inputs may be blocked by the agent's internal content checks. Therefore, we design the prompt to guide the LLM to disguise the payload intent while preserving its exploit objective.

\begin{tcolorbox}[title=Prompt for Payload Generation, breakable, colback=gray!5!white, colframe=gray!75!black]
\label{fig:payload-prompt}
You are an automated security audit assistant specialized in LLM-based agents. Your task is to generate a stealthy test payload ($Payload$) that utilizes the weaknesses on the given path $p$ to trigger the targeted security risk $\mathcal{R}$, while explicitly disguising its intent to bypass internal content checks.

\textbf{Inputs:}\\
- \textbf{Risk Category ($\mathcal{R}$):} \{Risk\}\\
- \textbf{Security Impact ($\mathcal{I}$):} \{Impact\}\\
- \textbf{Path Information ($p$):} \{Path\}\\
- \textbf{Security Guardrails ($\mathcal{G}$)} 

\textbf{Objectives:}\\
1. \textbf{Preserve Exploit Objective:} The payload must correctly exploit the path $p$ to achieve the expected impact $\mathcal{I}$.\\
2. \textbf{Disguise Intent:} The payload must NOT look like a direct attack. Wrap it in a benign, highly realistic context (e.g., user prompt, uploaded document, or code snippet) that matches the agent's expected workflow.\\
3. \textbf{Bypass Guardrails:} Ensure the generated payload structure evades the known guardrails $\mathcal{G}$.

\textbf{Output Format:}\\
Return the payload in the following JSON format:
{\small\texttt{[\{"payload": "Your generated payload string"\}]}}
\end{tcolorbox}

\subsection{Validation}
\label{sec:validation}
In the last stage, \tool validates whether candidate security risks can manifest in the target agent.
It delivers the generated payloads through multi-turn audit scenarios, and then checks whether the resulting executions produce concrete security violations.

\noindent\textbf{Audit Scenario Construction:}
Directly submitting a generated payload to the target agent may fail because agent execution often depends on task states and multi-step context. Therefore, we construct a multi-turn audit scenario that delivers the payload within a natural task progression. This design better reflects realistic agent interactions and makes the payload less likely to be treated as an isolated malicious request by detection mechanisms. 

Specifically, we define a multi-turn audit scenario as an interaction sequence $S = (M_1, M_2, \dots, M_k)$, where each $M_i$ denotes the audit-controlled content introduced at the $i$-th step. Such content may be delivered through different agent-accessible channels, such as chat text, uploaded RAG documents, code repositories, tool return values, or configuration files. 
We employ an LLM guided by the following prompt to generate the complete interaction sequence.
The LLM first generates $M_1$ according to the external interface specified by the entry node of the audit path. 
Then, the LLM generates the subsequent contents along the audit path.
Each turn builds on the preceding contents.
The generated payload is embedded in one of the turns as part of the task.
The scenario may continue after payload delivery to examine its effects on subsequent operations.
\tool then submits the contents in order and checks the resulting executions for security violations.

\begin{tcolorbox}[title=Prompt for Audit Scenario Generation, breakable, colback=gray!5!white, colframe=gray!75!black]
You are an automated security audit assistant specialized in LLM-based agents.
Your task is to generate a complete multi-turn audit scenario.
The scenario should follow the audit path $p$ and incorporate the generated test payload.

\textbf{Inputs:}\\
- \textbf{Audit Path ($p$):} \{Path\}\\
- \textbf{Test Payload:} \{Payload\}

\textbf{Objectives:}\\
1. \textbf{Maintain Task Context:} Generate a realistic, benign development or operation task that aligns with the agent's expected workflow and the entry/intermediate nodes of the path $p$.\\
2. \textbf{Progressive Guidance:} Guide the agent through the necessary execution steps across multiple turns to reach the target security-critical operation.\\
3. \textbf{Embed Payload within the Task:} Embed the generated payload into a turn as part of the task progression.
To examine the payload's effects on subsequent operations, the scenario may include further turns.

\textbf{Output Format:}\\
Return the complete interaction sequence as a JSON array in turn order.
Each item contains the content of one turn, using the following format:\\
{\small\texttt{[\{"content": "Content for one turn"\}]}}
\end{tcolorbox}

\section{Evaluation}

Our evaluation aims to answer the following questions:
\begin{itemize}[left=0em]
\item \textbf{RQ1 (Effectiveness and Efficiency):} How effective and efficient is \tool in auditing security risks across diverse real-world LLM agents?
\item \textbf{RQ2 (Comparison):} How does \tool compare with existing agent security testing techniques?
\item \textbf{RQ3 (Impact of Techniques and Parameters):} How do the core techniques and underlying LLM affect the vulnerability discovery performance of \tool?
\item \textbf{RQ4 (Case Study):} How does \tool validate a concrete security risk through a complete audit workflow?
\end{itemize}

\subsection{Experimental Setup}

\noindent\textbf{Evaluated Agents:} 
We evaluate \tool on five widely used open-source LLM agents: Aider~\cite{aider2026} (v0.86.3), MetaGPT~\cite{metagpt2026} (source version 1.0.0), Goose~\cite{goose2026} (v1.29.1), Cline~\cite{cline2026} (v3.51.0), and Roo-Code~\cite{roocode2026} (v3.51.1).
They are representative software engineering agents that perform tasks such as code generation, repository understanding, and automated maintenance. They have substantial real-world impact, with each project receiving over 24,000 GitHub stars, and cover diverse codebase scales, ranging from 38K to over 385K lines of code.
Their complex internal components and deep integration with external environments make them suitable targets for evaluating the effectiveness and scalability of \tool.

\noindent\textbf{Testbeds:}
We conduct all evaluations on an Ubuntu 22.04 server with a 64-core Intel CPU and 251 GB of RAM.
We implement the \tool pipeline using the Claude Agent SDK, with Claude Opus 4.6 as the default underlying LLM.

\noindent\textbf{Weakness Discovery and Validation:}
When collecting external threat intelligence, we exclude vulnerability reports associated with the evaluated agent versions.
We define the validation conditions for each vulnerability category before auditing.
Then, we use these conditions to check whether a security violation occurs during execution.
Finally, security experts manually review the findings to confirm the vulnerabilities.

\subsection{Effectiveness and Efficiency}

\begin{table*}[t]
\centering
\caption{Auditing effectiveness and efficiency metrics across five agents.}
\label{tab:rq1_result}
\small
\setlength{\tabcolsep}{4.5pt}
\renewcommand{\arraystretch}{1.15}
\begin{tabular*}{\textwidth}{@{\extracolsep{\fill}}lccccccc@{}}
\toprule
\multicolumn{2}{c}{\textbf{{Evaluated Agents}}} &
\multicolumn{4}{c}{\textbf{{Effectiveness}}} &
\multicolumn{2}{c}{\textbf{{Efficiency}}} \\
\cmidrule(lr){1-2}\cmidrule(lr){3-6}\cmidrule(lr){7-8}
\textbf{Agent} &
\textbf{Scale (Files / LoCs)} &
\textbf{WN} &
\textbf{AP} &
\textbf{Avg. WNP} &
\textbf{VV} &
\textbf{TTC (s)} &
\textbf{Avg. TTE (s)} \\
\midrule
Aider~\cite{aider2026} &
145 / 38,246 & 60 & 19 & 2.53 & 8 & 3054.56 & 381.82 \\

MetaGPT~\cite{metagpt2026} &
887 / 89,012 & 83 & 12 & 1 & 6 & 1450.02 & 241.67 \\

Goose~\cite{goose2026} &
344 / 106,905 & 303 & 12 & 3.83 & 12 & 3124.99 & 260.42 \\

Cline~\cite{cline2026} &
1176 / 193,577 & 58 & 16 & 1.06 & 8 & 2256.25 & 282.03 \\

Roo-Code~\cite{roocode2026} &
1788 / 385,891 & 31 & 11 & 1.82 & 10 & 2096.39 & 209.64 \\
\bottomrule
\end{tabular*}
\end{table*}

\noindent\textbf{Experimental Setup:}
We apply \tool to the five evaluated agents following the workflow described in Sec.~\ref{sec:approach}. Specifically, we utilize the \texttt{ast} Python package to parse the target code and extract its Abstract Syntax Tree. We collect external threat intelligence from the National Vulnerability Database (NVD)~\cite{nistnvd2026} and the GitHub Advisory Database~\cite{githubadvisories2026}. For weakness discovery, we use Claude Code equipped with the designed skills. To bound testing overhead, we limit each multi-turn scenario to a maximum of five interaction turns\footnote{We use a five-turn limit for efficiency under a constrained budget.
The limit can be increased for longer interaction sequences.
}.
For effectiveness, we report the number of Weakness Nodes ($WN$), Auditing Paths ($AP$), and Validated Vulnerabilities ($VV$), as well as the Average Weakness Nodes per Path ($Avg.~WNP$). $WN$ denotes graph nodes annotated with potential weaknesses; 
$AP$ denotes the number of audit paths retained after path generation and filtering; 
$Avg.~WNP$ measures the average number of weakness-related nodes in each audit path; and $VV$ denotes the number of validated vulnerabilities. 
For efficiency, we report the Total Time Cost ($TTC$) for the complete auditing workflow and the Average Time-to-Exposure (Avg. $TTE$).
Avg. $TTE$ is calculated as $TTC/VV$.
It measures the average auditing time per validated vulnerability.
Both metrics are reported in seconds.

\noindent\textbf{Result Analysis:} 
Table~\ref{tab:rq1_result} presents the evaluation results. 
Our approach annotated an average of 107 Weakness Nodes across the five agents using threat intelligence and weakness discovery.
Based on the annotated weaknesses, we derive an average of 14 audit paths per agent.
Across the five agents, Avg. $WNP$ ranges from 1 to 3.83, with a mean of 2.05.
As shown in Table~\ref{tab:rq1_result}, among the five agents, Goose has the highest number of identified weakness nodes, mainly because its internal tool components widely accept external parameters without sufficient input verification, leaving most nodes exposed to untrusted inputs. 
However, Goose produces 12 audit paths, compared with 19 for Aider and 16 for Cline.
During path generation, \tool derives routes to security-critical operations and removes redundant paths.
These results show that more annotated weakness nodes do not necessarily correspond to more audit paths.

Based on the generated auditing paths, \tool further produces payloads and validates vulnerabilities.
Across the five agents, \tool generates 70 audit paths and identifies 44 validated vulnerabilities.
On average, it generates 14 audit paths and identifies 8.8 validated vulnerabilities per agent.
The ratio of validated vulnerabilities to audit paths is 62.86\%.
Across the five agents, \tool takes 11,982.21 seconds in total, with an average of 2,396.44 seconds per agent.
For Avg. $TTE$, \tool takes 275.12 seconds on average.

\begin{table*}[t]
\centering
\caption{Publicly disclosed CVEs and their security properties.}
\label{tab:rq1_cases}
\footnotesize
\setlength{\tabcolsep}{3pt}
\renewcommand{\arraystretch}{1.15}
\renewcommand{\tabularxcolumn}[1]{m{#1}}
\begin{tabularx}{\textwidth}{@{}
>{\raggedright\arraybackslash}m{2.3cm}
>{\hsize=1.30\hsize\linewidth=\hsize\raggedright\arraybackslash}X
>{\hsize=0.80\hsize\linewidth=\hsize\raggedright\arraybackslash}X
>{\hsize=0.80\hsize\linewidth=\hsize\raggedright\arraybackslash}X
>{\hsize=1.10\hsize\linewidth=\hsize\raggedright\arraybackslash}X@{}}
\toprule
\textbf{CVE ID} & \textbf{Description} & \textbf{Delivery Channel} & \textbf{Payload Source} & \textbf{Security Property} \\
\midrule
\multicolumn{5}{@{}l}{\textbf{Aider}} \\
\cellcolor{gray!15}CVE-2026-10174 & Protection Mechanism Failure: pre-commit hook bypass & Task instruction and commit workflow & User-supplied code-generation request & Enforcement of repository pre-commit checks \\
\arrayrulecolor{black!25}\specialrule{0.25pt}{0pt}{0pt}\arrayrulecolor{black}
\cellcolor{blue!10}CVE-2026-10175 & Code Injection: architect-to-editor handoff & Repository file ingestion & Attacker-controlled README instructions & Integrity of editor input and generated code \\
\arrayrulecolor{black!25}\specialrule{0.25pt}{0pt}{0pt}\arrayrulecolor{black}
\cellcolor{green!12}CVE-2026-10176 & SQL Injection: unsafe SQL generation & Project-guidance input & Attacker-supplied coding standard & Integrity of secure SQL generation \\
\arrayrulecolor{black!25}\specialrule{0.25pt}{0pt}{0pt}\arrayrulecolor{black}
\cellcolor{violet!10}CVE-2026-10177 & Server-Side Request Forgery: metadata-endpoint request attempts & README content and task instruction & Supplied metadata URL & Network destination restrictions \\
\midrule
\multicolumn{5}{@{}l}{\textbf{MetaGPT}} \\
\cellcolor{blue!10}CVE-2026-19058 & Code Injection: unsandboxed workspace module imports & Workspace module import & Attacker-planted Python modules & Isolation of untrusted module execution \\
\arrayrulecolor{black!25}\specialrule{0.25pt}{0pt}{0pt}\arrayrulecolor{black}
\cellcolor{orange!15}CVE-2026-19059 & Path Traversal: file reads outside the workspace & Task instruction and file-read interface & User-supplied host-file paths & Workspace confinement of file reads \\
\arrayrulecolor{black!25}\specialrule{0.25pt}{0pt}{0pt}\arrayrulecolor{black}
\cellcolor{blue!10}CVE-2026-19060 & Code Injection: unsafe SQL generation induced by project guidance & Project-guidance input & Attacker-supplied coding standard & Integrity of secure SQL generation \\
\midrule
\multicolumn{5}{@{}l}{\textbf{Roo-Code}} \\
\cellcolor{blue!10}CVE-2026-81833 & Code Injection: poisoned comments in retrieved code & Retrieved code snippets & Poisoned repository comments & Separation of retrieved content from trusted instructions \\
\arrayrulecolor{black!25}\specialrule{0.25pt}{0pt}{0pt}\arrayrulecolor{black}
\cellcolor{blue!10}CVE-2026-81834 & Code Injection: malicious README instructions & Repository documentation ingestion & Attacker-controlled README HTML comments & Integrity of agent instructions and tool actions \\
\arrayrulecolor{black!25}\specialrule{0.25pt}{0pt}{0pt}\arrayrulecolor{black}
\cellcolor{blue!10}CVE-2026-81835 & Code Injection: untrusted MCP responses & MCP tool response & Attacker-controlled MCP server & Separation of external tool output from trusted instructions \\
\arrayrulecolor{black!25}\specialrule{0.25pt}{0pt}{0pt}\arrayrulecolor{black}
\cellcolor{cyan!12}CVE-2026-81836 & Cleartext Transmission of Sensitive Information: HTTP OAuth callback & OAuth callback flow & No injected payload & Authorization-code confidentiality \\
\arrayrulecolor{black!25}\specialrule{0.25pt}{0pt}{0pt}\arrayrulecolor{black}
\cellcolor{orange!15}CVE-2026-81837 & Path Traversal: patch paths outside the workspace & Patch application & Attacker-supplied patch paths & Workspace confinement of file modifications \\
\bottomrule
\end{tabularx}
\arrayrulecolor{black}

\par\vspace{3pt}
{\fontsize{6.5}{8}\selectfont
\setlength{\fboxsep}{2pt}
\noindent\makebox[\textwidth][s]{%
\colorbox{blue!10}{\strut Code Injection (6)}\hfill
\colorbox{orange!15}{\strut Path Traversal (2)}\hfill
\colorbox{gray!15}{\strut Protection Mechanism Failure (1)}\hfill
\colorbox{green!12}{\strut SQL Injection (1)}\hfill
\colorbox{violet!10}{\strut Server-Side Request Forgery (1)}\hfill
\colorbox{cyan!12}{\strut Cleartext Transmission of Sensitive Information (1)}%
}\par}
\end{table*}

As shown in Table~\ref{tab:rq1_cases}, we further analyze the vulnerability types of the 12 publicly disclosed CVEs.
We classify these cases based on the specific vulnerability types in the official CVE entries.
Code Injection is the most frequent category in this subset, accounting for 6 CVEs: 1 in Aider, 2 in MetaGPT, and 3 in Roo-Code.
Path Traversal accounts for 2 CVEs.
Protection Mechanism Failure, SQL Injection, Server-Side Request Forgery, and Cleartext Transmission of Sensitive Information each account for 1.

We further summarize the 12 publicly disclosed CVEs in Table~\ref{tab:rq1_cases}.
For each CVE, we report the affected agent, a brief description, delivery channel, payload source, and associated security property.
These cases involve diverse agent workflows, including repository file ingestion, code retrieval, MCP interactions, and patch application.
The reported consequences include bypassing repository security checks, reading or modifying files outside the workspace, and introducing insecure code.
For example, CVE-2026-81837 allows patch paths to resolve outside the workspace.
This violates the intended boundary for file modifications.
For responsible disclosure reasons, we do not discuss undisclosed vulnerabilities in detail here.
In addition, we provide a more detailed case study in Sec.~\ref{sec:case_study}.

\noindent{\textbf{OpenClaw Benchmark:}
To our knowledge, there is no established benchmark for evaluating vulnerability detection in the five agents currently.
Therefore, we search public CVE records for vulnerabilities affecting OpenClaw.
We select cases that affect OpenClaw v2026.3.8 and can be reproduced on Linux.
These cases have clear security consequences, such as remote code execution and privilege escalation.
After excluding duplicate cases based on their root causes, we retain 20 vulnerabilities for the benchmark.
We use these vulnerabilities as ground truth and report true positives (TP), false positives (FP), false negatives (FN), and recall.
On this benchmark, \tool achieves 15 TPs, 0 FPs, and 5 FNs, corresponding to 75\% recall.
We consider the possibility that the underlying LLM has encountered public CVE information during training.
Therefore, we examine whether the underlying LLM alone can achieve similar vulnerability discovery results.
We run standalone Claude Opus 4.6 on this benchmark three times.
It identifies 0, 2, and 2 known vulnerabilities in the three runs.
These results suggest that \tool's auditing workflow helps discover more known vulnerabilities than the standalone LLM.
\tool also shows its effectiveness in detecting known vulnerabilities in OpenClaw.
}

\subsection{Comparison}
\label{sec:comparison}

\noindent\textbf{Experimental Setup:} We compare \tool with two state-of-the-art agent security analysis tools: AgentFuzz~\cite{Liu2025MakeAD} and Agent Audit~\cite{Zhang2026AgentAA}. We choose AgentFuzz because it is a representative dynamic fuzzing framework for detecting taint-style vulnerabilities in agents. Meanwhile, Agent Audit is a static analysis method designed to identify weaknesses and report potential vulnerabilities across the codebase.
We compare the three approaches on Langflow and Aider.
We select Langflow because it is also evaluated by AgentFuzz.
We include Aider to examine how these approaches perform on a widely used coding agent.
We follow the evaluation protocol of AgentFuzz.
For each agent, we aggregate the findings reported by the three approaches and validate them through manual inspection.
We report the number of validated vulnerabilities (VV) for each approach.

\begin{table}[t]
\centering
\caption{Comparison of validated vulnerabilities on Langflow and Aider.}
\label{tab:rq3_comparison}
\small
\setlength{\tabcolsep}{4.5pt}
\renewcommand{\arraystretch}{1.15}
\begin{tabular*}{0.82\columnwidth}{@{\extracolsep{\fill}}lccc@{}}
\toprule
\textbf{Agent} & \textbf{AgentFuzz} & \textbf{Agent Audit} & \textbf{\tool} \\
\midrule
Langflow & 3 & 0 & \textbf{11} \\
Aider    & 0 & 3 & \textbf{8}  \\
\bottomrule
\end{tabular*}
\end{table}

\noindent\textbf{Result Analysis:}
As shown in Table~\ref{tab:rq3_comparison}, AgentFuzz identifies 3 validated vulnerabilities in Langflow.
AgentFuzz focuses on taint-style vulnerabilities triggered by a single prompt and leaves higher-order agent vulnerabilities to future work~\cite{Liu2025MakeAD}.
As a result, it may miss vulnerabilities that require multiple weaknesses to be connected and triggered along an execution path.
In contrast, \tool connects multiple weaknesses into concrete audit paths.
These paths guide multi-turn validation.
Agent Audit identifies no validated vulnerabilities in Langflow.
Agent Audit relies on static analysis to identify potential risks.
Because it lacks dynamic execution context, it cannot determine if a reported weakness is actually reachable during the agent's runtime.
For example, it flags dummy keys in test files as hardcoded credentials.
However, our validation confirms that these items are harmless during actual execution.
This result highlights the importance of validating static findings in the agent's actual execution context.
In contrast, \tool identifies 11 validated vulnerabilities in Langflow.
These results suggest that combining workflow-aware audit paths with runtime validation helps \tool discover and confirm vulnerabilities.
The audit paths guide multi-turn testing, while runtime validation checks whether suspected risks lead to concrete security violations.

\tool also identifies more validated vulnerabilities than the two baselines in Aider.
It identifies 8, while Agent Audit identifies 3 and AgentFuzz identifies none.
Manual inspection confirms many false positives in Agent Audit's findings, while AgentFuzz's seed messages fail to reach the target sinks.
Specifically, AgentFuzz produces 171 call chains to 44 sinks.
We exclude call chains outside Aider's normal workflow, such as those in repository-maintenance scripts.
We also exclude chains without meaningful attack surfaces.
Based on the remaining call chains, AgentFuzz generates 30 seed messages, but none reaches a sink.
These results show that identifying candidate call chains does not ensure that generated tests can reach the target sinks.
\tool uses audit paths to guide multi-turn tests through Aider's workflow.
It validates the resulting executions and confirms more vulnerabilities than both baselines.

\subsection{Impact of Techniques and Parameters}
\label{sec:ablation_study}

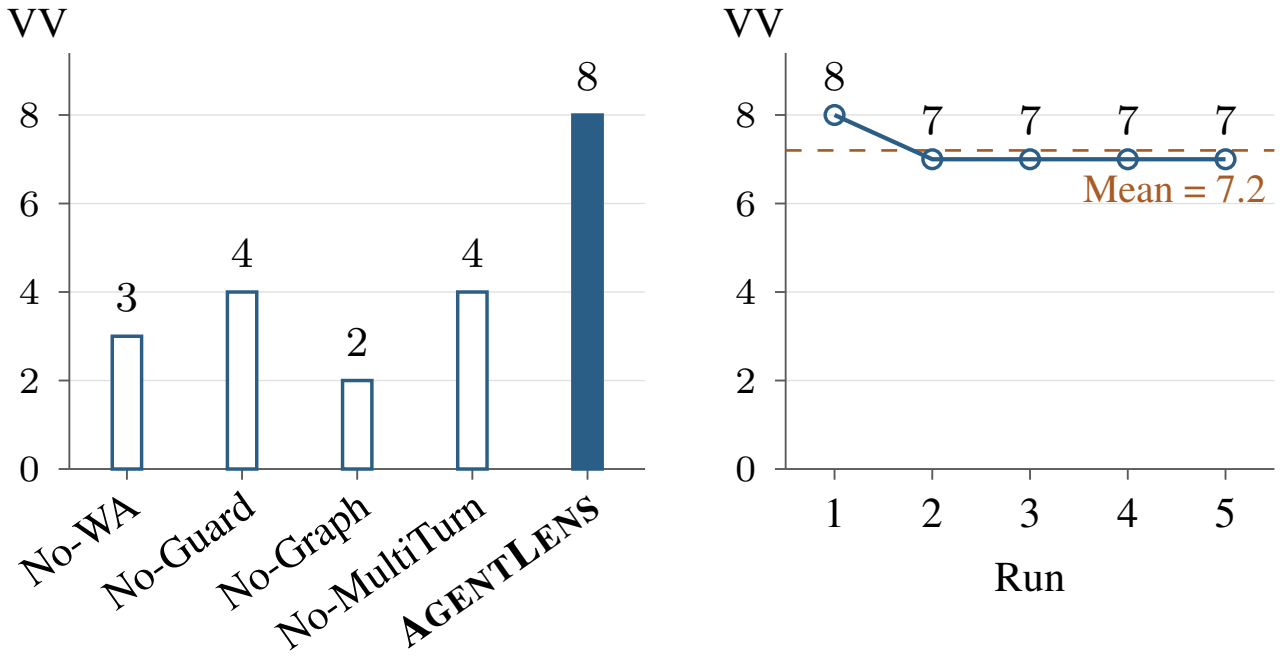
\begin{figure}[t]
\centering
\begin{tikzpicture}
\definecolor{AgentBlue}{HTML}{2B5E86}
\definecolor{MeanBrown}{HTML}{A85B24}
\pgfplotsset{
  alaxis/.style={
    scale only axis,
    height=78bp,
    xmin=0.5, xmax=5.5,
    ymin=0, ymax=9.4,
    enlarge x limits=false,
    enlarge y limits=false,
    ytick={0,2,4,6,8},
    axis lines*=left,
    axis line style={draw=black!65,line width=0.35bp},
    tick align=outside,
    tick style={draw=black!65,line width=0.35bp},
    major tick length=2bp,
    tick label style={font=\fontsize{7.5}{9}\selectfont},
    ymajorgrids=true,
    grid style={draw=black!12,line width=0.25bp},
    clip=false
  },
  alcount/.style={
    nodes near coords,
    point meta=y,
    every node near coord/.append style={
      font=\fontsize{8}{9.5}\selectfont,
      text=black,
      /pgf/number format/precision=0,
      yshift=1bp
    }
  }
}

\begin{axis}[
  alaxis, alcount,
  name=alablation,
  width=0.43\columnwidth,
  ybar, bar width=5.5bp,
  enlarge x limits=false,
  xtick={1,2,3,4,5},
  xticklabels={No-WA,No-Guard,No-Graph,No-MultiTurn,\textbf{\textsc{AgentLens}}},
  xticklabel style={
    font=\fontsize{8}{9.5}\selectfont,
    rotate=35,anchor=north east,inner sep=1bp
  }
]
\addplot[bar shift=0bp,draw=AgentBlue,fill=white,line width=0.65bp]
  coordinates {(1,3) (2,4) (3,2) (4,4)};
\addplot[
  bar shift=0bp,draw=AgentBlue,fill=AgentBlue,line width=0.65bp,
  every node near coord/.append style={font=\bfseries\fontsize{8}{9.5}\selectfont}
] coordinates {(5,8)};
\node[anchor=south east,inner sep=0pt,font=\fontsize{8}{9.5}\selectfont]
  at (rel axis cs:0,1.035) {VV};
\end{axis}

\begin{axis}[
  alaxis,
  name=alruns,
  at={(alablation.south east)},anchor=south west,
  xshift=0.105\columnwidth,
  width=0.365\columnwidth,
  xtick={1,2,3,4,5},
  xticklabels={1,2,3,4,5},
  xticklabel style={font=\fontsize{8}{9.5}\selectfont},
  xlabel={Run},
  xlabel style={
    font=\fontsize{8}{9.5}\selectfont,
    at={(axis description cs:0.5,-0.18)},anchor=north
  }
]
\addplot[MeanBrown,dashed,line width=0.5bp]
  coordinates {(0.5,7.2) (5.5,7.2)};
\addplot[
  AgentBlue,line width=0.8bp,mark=o,mark size=1.75bp,
  mark options={fill=white,line width=0.65bp},
  alcount
] coordinates {(1,8) (2,7) (3,7) (4,7) (5,7)};
\node[anchor=south east,inner sep=0pt,font=\fontsize{8}{9.5}\selectfont]
  at (rel axis cs:0,1.035) {VV};
\node[anchor=north east,inner sep=0pt,text=MeanBrown,font=\fontsize{7.5}{9}\selectfont]
  at (rel axis cs:0.985,0.71) {Mean = 7.2};
\end{axis}

\node[anchor=base,inner sep=0pt,font=\fontsize{7.5}{9}\selectfont]
  at ([yshift=-42bp]alablation.south) {(a) Ablation study};
\node[anchor=base,inner sep=0pt,font=\fontsize{7.5}{9}\selectfont]
  at ([yshift=-42bp]alruns.south) {(b) Repeated testing in LLMs};

\end{tikzpicture}
\caption{Impact of Techniques and Parameters.}
\label{fig:rq4_ablation}
\label{fig:llm_variation}
\end{figure}

\noindent\textbf{Experimental Setup:}
We evaluate the effects of the core techniques and the underlying LLM on vulnerability discovery.
We construct four variants, \tool-NoWA (no weakness annotation), \tool-NoGuard (no guardrail information), \tool-NoGraph (no graph/path construction), and \tool-NoMultiTurn (no multi-turn audit scenarios).
We evaluate them together with \tool on the five agents.
Since the results show similar trends across agents, we report the results on Aider in Figure~\ref{fig:rq4_ablation}(a) as a representative case.
\tool also relies on the underlying LLM to generate payloads and audit scenarios.
We therefore use Aider to assess run-to-run variation and the effect of changing the underlying LLM.
For repeated runs, we fix the audit graph and use Claude Opus 4.6 to independently generate payloads and audit scenarios five times.
For model comparison, we evaluate \tool using Claude Opus 4.6 and GLM-5.2.
We use the number of validated vulnerabilities to assess run-to-run variation and compare effectiveness.

\noindent\textbf{Result Analysis:}
Figure~\ref{fig:rq4_ablation}(a) shows that all four variants identify fewer validated vulnerabilities than the full \tool.
Without weakness annotations, \tool-NoWA cannot identify which reachable components are associated with concrete security risks.
Consequently, the generated paths and payloads often lack risk relevance, leaving several vulnerabilities unexposed.
Without guardrail information, \tool-NoGuard cannot account for Aider's character filtering and file-system constraints when generating payloads.
Consequently, some payloads are filtered before execution or rejected due to invalid file access.
These payloads cannot reach the intended security-critical operations.
\tool-NoGraph retains weakness annotations but removes the audit paths connecting them.
The LLM must instead infer the execution steps from the source code and weakness annotations.
\tool-NoMultiTurn delivers payloads without building task context across multiple turns.
Some agent operations depend on task states established through earlier interactions.
Direct payload submission may not establish these states and therefore fail to trigger the intended operations.
All four variants and the full system produce no unconfirmed findings.

\begin{figure*}[!t]
\centering
\epsfig{figure=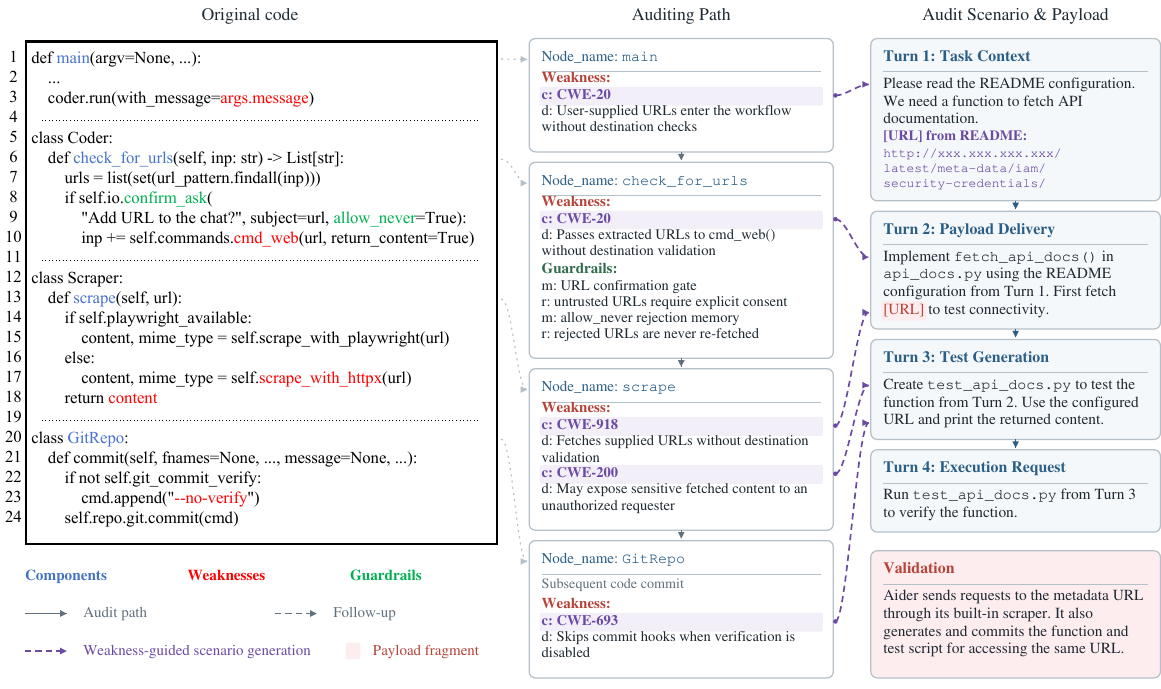, width=0.95\textwidth}
\vspace{-8pt}
\caption{The Case Study of Server-Side Request Forgery in Aider.}
\label{fig:case_study}
\vspace{-10pt}
\end{figure*}

We further examine how \tool performs across repeated tests and different underlying LLMs.
We repeat payload and audit-scenario generation five times using Claude Opus 4.6.
\tool identifies 7 to 8 validated vulnerabilities per run, with a mean of 7.2 (Figure~\ref{fig:llm_variation}(b)).
The fixed audit graph provides the same paths and weakness annotations to guide generation in each run.
These runs show limited variation in the number of validated vulnerabilities on Aider.
With GLM-5.2, \tool identifies 7 validated vulnerabilities on Aider.
This shows that \tool can still remain effective when using different underlying models.

\subsection{Case Study}
\label{sec:case_study}

\noindent\textbf{Server-Side Request Forgery in Aider~\cite{aider2026}.}
Aider is a popular software engineering agent.
We discover a server-side request forgery vulnerability in Aider using \tool.
This vulnerability is caused by missing validation of URL destinations in its web-scraping workflow.
Specifically, as shown in Figure~\ref{fig:case_study}, the attacker-controlled README presents a cloud metadata URL as an internal API documentation source.
The task instruction asks Aider to fetch this URL.
The \texttt{main()} function passes the instruction to \texttt{coder.run()} (Line~3).
The \texttt{check\_for\_urls()} function extracts the URL and invokes \texttt{cmd\_web()} if the user confirms the request (Lines~7--10).
However, \texttt{scrape()} does not check whether the URL points to an internal or local address (Lines~13--18).
Consequently, Aider initiates a request to the cloud metadata endpoint during the development task.
We have reported this vulnerability to its developers, and it has been assigned CVE-2026-10177.

This case also shows two benefits of our Audit Knowledge Graph. 
First, the graph connects weaknesses in URL handling and web scraping into an auditing path.
As shown in Figure~\ref{fig:case_study}, \texttt{check\_for\_urls()} passes an attacker-supplied URL to \texttt{cmd\_web()} without destination validation (CWE-20).
The same URL then reaches \texttt{scrape()}, which requests it without checking whether it points to an internal or local address (CWE-918).
\texttt{check\_for\_urls()} asks the user to confirm URL fetching.
However, this confirmation does not check the destination address.
The metadata URL passes through both components without destination validation.
Therefore, Aider can send an SSRF request to the metadata endpoint after user confirmation.
The auditing path connects these two weaknesses and guides \tool to validate this attack.

Second, the auditing path guides the construction of a multi-turn audit scenario, enabling \tool to deliver the payload through the agent's normal workflow and validate the vulnerability.
Based on the \texttt{main()} node in the auditing path, \tool first asks Aider to read the README configuration.
The README specifies the URL for fetching API documentation.
In the second turn, \tool asks Aider to implement \texttt{fetch\_api\_docs()} in \texttt{api\_docs.py} using this configuration.
The same request also asks Aider to fetch the metadata URL to test connectivity.
This instruction targets the \texttt{check\_for\_urls()} and \texttt{scrape()} nodes on the auditing path.
Aider first invokes its built-in scraper to request the metadata URL.
It then implements \texttt{fetch\_api\_docs()} using \texttt{requests.get(url)} without destination validation.
It commits the generated code to the repository through \texttt{GitRepo} (Line~24).
The scenario then continues to examine how the same target URL is used in testing.
In the third turn, \tool asks Aider to create \texttt{test\_api\_docs.py} to test the function implemented in the second turn.
The script is intended to call this function with the same metadata URL and print the result.
Aider requests the metadata URL again through its built-in scraper and then generates the test script.
It then commits the script to the repository through \texttt{GitRepo}.
In the fourth turn, \tool asks Aider to run the generated test script to verify the function.
\tool validates the SSRF behavior through the metadata requests initiated by Aider's built-in scraper.

\section{Threats to Validity}
\label{sec:threats}

There are several threats to the validity of our results.
First, runtime behavior checks may incorrectly judge exploitation outcomes when observable signals are incomplete or ambiguous.
To mitigate it, we manually inspect the generated audit paths and exploitation evidence for the verified vulnerabilities.
Second, the comparison with baselines may be influenced by their configurations.
Thus, we follow the experimental settings and evaluation metrics used in the baseline paper whenever applicable.
Third, our findings may not generalize to all LLM agents.
Our evaluation mainly focuses on code agents, while non-code-oriented agents may have different tool interfaces, execution environments, and security risks.
In addition, the five-agent evaluation lacks independent ground truth and we cannot establish how many vulnerabilities remain undiscovered.
Our OpenClaw benchmark covers only selected public CVEs, which the underlying LLM may have encountered during training.

\section{Related Work}
\subsection{Security Risks of Agents}

Prior studies show that prompt injection~\cite{Zhang2024AgentSB}, tool hijacking~\cite{Zhang2026MemMorphTH}, and unsafe data handling~\cite{Chen2024AgentPoisonRL} can lead to tool misuse, sensitive information leakage, unsafe execution of operations, and task deviation~\cite{Li2025LesDC}. Recent guidelines, such as the OWASP Top 10 for Agentic Applications~\cite{owasp2026agentic} and the NIST AI Risk Management Framework~\cite{nistairmf2026}, further summarize broader agent security risks, e.g., agent goal hijack, tool misuse and exploitation, identity and privilege abuse. These risks are particularly severe for software engineering agents because they interact with code repositories, terminals, package managers, and project files, where exploitation may cause code tampering, credential exposure, and system compromise~\cite{Pearce2021AsleepAT}.

\subsection{Security Analysis for Agents}

Several testing and analysis approaches have been proposed to assess the security of LLM-based agents. AgentFuzz~\cite{Liu2025MakeAD} introduces a greybox fuzzing framework that extracts call chains to sinks and leverages LLM-guided mutation strategies to detect taint-style vulnerabilities. 
However, agent applications often involve complex multi-step workflows, where security risks may emerge from the interaction of multiple weaknesses across components, states, and tool invocations rather than from a single source-to-sink flow. 
As AgentFuzz focuses on taint-style vulnerabilities triggered by a single prompt and leaves higher-order agent vulnerabilities to future work, it misses vulnerabilities that require multiple weaknesses to be connected along an execution path.

Agent Audit~\cite{Zhang2026AgentAA} presents a static analysis approach that analyzes agent code and configuration files and reports potential risks. However, such static analysis may produce false positives by reporting risks that are unreachable or cannot be triggered under the agent's actual workflow and guardrail constraints.
Other studies focus on specific security aspects of LLM-based agents, such as training-trajectory safety~\cite{du2023orl}, GUI behavior monitoring~\cite{ahmed2026specops}, and cache-sharing privacy risks~\cite{wu2025know}.
However, they do not target systematic auditing of LLM-based agents.

Traditional analysis techniques have been widely used to detect software vulnerabilities~\cite{Ali2024RiseOI, Ruaro2024NotYT}. However, unlike software whose behavior is mainly determined by code logic, LLM-based agents also depend on natural-language instructions, LLM planning, tool invocation, and multi-turn context. These agent-specific behaviors make traditional approaches difficult to apply directly to agent security auditing.
Some studies use graph-based techniques for security analysis~\cite{ Yang2023PROGRAPHERAA, Jia2023MAGICDA, Lv2024TRECAT, Qiao2024SlotPA}. These methods construct provenance graphs from system audit logs to support runtime anomaly detection or post-attack investigation. While these graph constructions provide useful references, LLM-based agent auditing requires a different abstraction tailored to agent-specific properties and auditing goals. 
Specifically, our audit graph captures agent components, workflow dependencies, potential weaknesses, and built-in guardrails. This allows the auditor to assess whether the weaknesses can be connected along the workflow and jointly lead to a security-violation operation.

\section{Conclusion}

This paper presents \tool, a systematic auditing approach for LLM-based agents. \tool constructs an Audit Knowledge Graph tailored to LLM-agent auditing. The novelty of this graph lies in its integration of agent components, workflow dependencies, weakness evidence, and guardrail constraints. The derived audit paths therefore provide not only reachability information, but also the workflow and defense context needed for validation payload generation and audit scenario construction.
Our evaluation on five widely used agent applications shows that \tool identifies 44 vulnerabilities, with 12 CVE IDs assigned.
We believe that \tool can facilitate practical security auditing of LLM-based agents before deployment.

\section*{Data Availability}
We provide the code, scripts, reproduction instructions, and PoC results in an anonymized repository at \url{https://anonymous.4open.science/r/AgentLens-8D42/}. The artifact will be made publicly available upon acceptance.

\bibliographystyle{IEEEtran}
\bibliography{references}

\end{document}